\documentclass[aps,pre,twocolumn,showpacs]{revtex4-1}
\usepackage{graphicx}  
\usepackage{subfigure}  
\usepackage{bm}        
\usepackage{amssymb}   
\usepackage[tbtags]{amsmath}

\usepackage{exscale}

\usepackage{xcolor}

\newcommand{\ket}[1]{|#1\rangle}

\newcommand{\bra}[1]{\langle#1|}

\definecolor{darkred}{rgb}{0.90,0.2,0.2}

\begin{document}

\title{Numerical linked cluster expansions for quantum quenches in one dimensional lattices}

\author{Krishnanand Mallayya}

\affiliation{Department of Physics, Pennsylvania State University,
University Park, PA, USA 16802}
\author{Marcos Rigol}
\affiliation{Department of Physics, Pennsylvania State University,
University Park, PA, USA 16802}


\begin{abstract}
We discuss the application of numerical linked cluster expansions (NLCEs) to study one dimensional lattice systems in thermal equilibrium and after quantum quenches from thermal equilibrium states. For the former, we calculate observables in the grand canonical ensemble, and for the latter we calculate observables in the diagonal ensemble. When converged, NLCEs provide results in the thermodynamic limit.  We use two different NLCEs - a maximally connected expansion introduced in previous works and a site-based expansion. We compare the effectiveness of both NLCEs. The site-based NLCE is found to work best for systems in thermal equilibrium. However, in thermal equilibrium and after quantum quenches, the site-based NLCE can diverge when the maximally connected one converges. We relate this divergence to the exponentially large number of clusters in the site-based NLCE and the behavior of the weights of observables in those clusters. We discuss the effectiveness of resummations to cure the divergence. Our NLCE calculations are compared to exact diagonalization ones in lattices with periodic boundary conditions. NLCEs are found to outperform exact diagonalization in periodic systems for all quantities studied.
\end{abstract}

\maketitle

\section{Introduction}
The study of strongly interacting lattice systems is in general very challenging and only special (integrable) models admit exact analytical solutions \cite{korepin_bogoliubov_book_93, sutherland_book_04, baxter2007exactly, cazalilla_citro_review_11}. In this context, numerical calculations based on series expansions have proved to be very useful \cite{domb1972phase, guttmann_book_89, oitmaa2006series}. An even more challenging question that has attracted much attention recently is the far-from-equilibrium quantum dynamics of isolated many-particle lattice systems \cite{dalessio_kafri_16, eisert_friesdorf_review_15, polkovnikov2011colloquium}. This has been motivated by recent experiments with ultracold atoms in optical lattices, in which it is possible to create strongly interacting quantum lattice systems with tunable parameters in nearly isolated environments. The far-from-equilibrium dynamics can be generated, e.g., by sudden changes of the depth of the optical lattice \cite{greiner2002collapse, will2010time, will2015observation} or by engineering special initial states~\cite{kinoshita2006quantum, gring2012relaxation, trotzky_chen_12, langen_erne_15, clos_porras_16, kaufman_tai_16}. A question that has been addressed in the latter set of experiments is whether, under (nearly) unitary dynamics, observables relax to time independent values that can be described using traditional statistical mechanics. 

The same question has been studied theoretically mostly for the quantum dynamics of pure states \cite{dalessio_kafri_16, eisert_friesdorf_review_15, polkovnikov2011colloquium}. Results from numerical simulations in a variety of nonintegrable lattice models in one and two dimensions have indicated that, under unitary dynamics, few-body observables relax to time-independent values that are described using traditional ensembles of statistical mechanics \cite{rigol2008thermalization, rigol2009breakdown, *rigol2009quantum, eckstein_kollar_09, rigol_santos_10, banuls_cirac_11, khatami_pupillo_13, zangara_dente_13, sorg_vidmar_14}, a phenomenon we call thermalization and which has been the central topic of a recent review~\cite{dalessio_kafri_16}. On the other hand, few-body observables in integrable systems relax to time-independent values that are described by the generalized Gibbs ensemble \cite{rigol2007relaxation,cazalilla_06, cassidy2011generalized, calabrese11, Ilievski15}, a phenomenon we call generalized thermalization and which has been a central topic of several recent reviews~\cite{vidmar_rigol_16, essler_fagotti_16, cazalilla_chung_16, caux_16, ilievski_medenjak_16}. Thermalization in nonintegrable systems has been understood to be the result of eigenstate thermalization \cite{srednicki1994chaos, deutsch1991quantum, rigol2008thermalization, rigol_srednicki_12, dalessio_kafri_16}, while generalized thermalization in integrable systems has been understood to be the result of generalized eigenstate thermalization \cite{cassidy2011generalized, vidmar_rigol_16}.

While integrable models can be studied using exact analytic approaches~\cite{vidmar_rigol_16, essler_fagotti_16, cazalilla_chung_16, caux_16, ilievski_medenjak_16}, nonintegrable ones require the use of numerical techniques such as full exact diagonalization \cite{rigol2008thermalization, rigol2009breakdown, *rigol2009quantum, rigol_santos_10, khatami_pupillo_13, zangara_dente_13, sorg_vidmar_14}, for which one can access arbitrarily long times but is limited to small system sizes, or density matrix renormalization group \cite{schollwock2005density,schollwock2011density} and dynamical mean field theory \cite{georges1996dynamical, Aoki14}  like approaches, for which the limitation is not in the system size but in the accessible times  \cite{eckstein_kollar_09, banuls_cirac_11, sorg_vidmar_14}. Recently, a numerical linked cluster expansion (NLCE) approach was introduced that, when converged, allows one to calculate infinite-time averages of observables in the thermodynamic limit after quenches from initial thermal equilibrium~\cite{rigol2014quantum} and pure \cite{rigol2014quantum2} states. This approach was used to probe fundamental differences in quenches to and away from integrability starting from thermal states~\cite{rigol2016fundamental}, and to study quantum quenches from N\'eel and tilted ferromagnetic states in the spin-1/2 XXZ chain~\cite{wouters_denardis_14_93, piroli_vernier_16}. Its main limitation is that the expansion may fail to converge when the temperature of the initial state is low~\cite{rigol2014quantum}, as well as for Hamiltonian parameters of interest in quenches from pure states~\cite{wouters_denardis_14_93, rigol2014quantum2, piroli_vernier_16}. For systems in thermal equilibrium, it has been shown that NLCEs converge to lower temperatures than traditional high-temperature expansions \cite{rigol2006numerical, *rigol2007numerical1, *rigol2007numerical2}, and exponentially faster than exact diagonalization calculations  \cite{iyer2015optimization}. NLCEs can also outperform other computational approaches in ground state \cite{khatami_singh_11, yang_schmidt_11, coester_clever_15, ixert_tischler_15} and entanglement entropy \cite{kallin_hyatt_13, stoudenmire_gustainis_14, sherman_devakul_16} calculations.

Here, we compare two complementary NLCEs, one based on maximally connected clusters (used in previous works \cite{rigol2014quantum, rigol2016fundamental}) and the second one is a site-based expansion. We compare the effectiveness of both expansions in one dimensional (1D) lattices with nearest and next-nearest neighbor couplings. Furthermore, we compare the NLCE results with exact diagonalization calculations in the grand canonical ensemble in finite systems with periodic boundary conditions. We consider systems in thermal equilibrium as well as infinite-time averages after quenches from thermal states. We show that NLCEs are superior to plain exact diagonalization. When the NLCEs fail to converge, we use resummation techniques to extend their region of convergence.

The presentation is organized as follows. In Sec.~\ref{sec:Ham}, we introduce the model Hamiltonian considered. The NLCEs used to study thermal equilibrium states and quantum quenches are discussed in Sec.~\ref{NLCE_sec}. There, we also review resummation methods used to accelerate the convergence, which were previously discussed in Ref.~\cite{rigol2006numerical, *rigol2007numerical1, *rigol2007numerical2} in the context of two-dimensional lattice systems in thermal equilibrium. The results of our calculation in 1D are presented in Sec.~\ref{GE_sec} for thermal states and in Sec.~\ref{DE_sec} for quantum quenches. In Sec.~\ref{weight_sec}, we discuss why the site-based NLCE fails to converge after quantum quenches for initial temperatures at which the thermal equilibrium results converge. A summary of our results is presented in Sec.~\ref{conclusion_sec}. 

\section{Model Hamiltonian}\label{sec:Ham}

We consider the $t$-$V$-$t'$-$V'$ model for hard-core bosons in 1D lattices~\cite{cazalilla_citro_review_11}
{\setlength\arraycolsep{0.5pt}
\begin{eqnarray}
\hat{H}&=&\sum_i \left\lbrace -t\left( \hat{b}^\dagger_i \hat{b}^{}_{i+1} + 
\textrm{H.c.} \right) 
+V\left( \hat{n}^{}_i-\dfrac{1}{2}\right)\left( \hat{n}^{}_{i+1}-\dfrac{1}{2}\right) 
\right.\nonumber\\
&-&\left.t'\left( \hat{b}^\dagger_i \hat{b}^{}_{i+2} + \textrm{H.c.} \right)  
+V'\left( \hat{n}^{}_i-\dfrac{1}{2}\right)\left( \hat{n}^{}_{i+2}-\dfrac{1}{2}\right)
\right\rbrace,\label{Hamilt}
\end{eqnarray}}\\
where $\hat{b}^\dag_i(\hat{b}^{}_i)$ is the hard-core boson creation (annihilation) operator at site $i$, and $\hat{n}^{}_i=\hat{b}^\dag_i\hat{b}^{}_i$ is the number operator. The creation-annihilation operators obey bosonic commutation relations: $[\hat b_i,\hat b_j]=0$, $[\hat b^\dag_i,\hat b^\dag_j]=0$, $[\hat b_i,\hat b_j^{\dagger}]=\delta_{ij}$, with the constraints $\hat b_i^2=(\hat b_i^{\dagger})^2=0$, which prevent multiple occupation of the lattice sites. Note that hopping and interaction terms are restricted to nearest $(t,V)$ and next-nearest $(t',V')$ neighbors sites. 

\begin{figure}[!b]
\centering
\includegraphics[width=70mm]{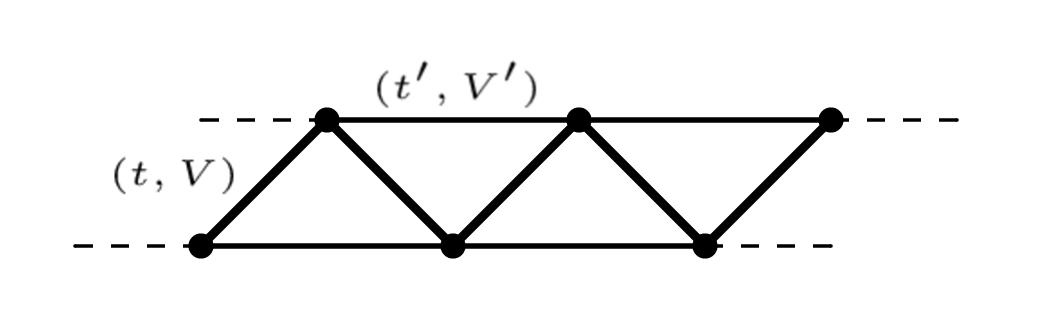}
\caption{1D lattice with nearest neighbor hoppings ($t$) and interactions $(V)$, and next-nearest neighbor hoppings ($t'$) and interactions $(V')$.}\label{1dlattice}
\end{figure}

This 1D lattice system can be represented graphically as depicted in Fig.~\ref{1dlattice}. In the absence of next-nearest neighbor hoppings and interactions, the Hamiltonian~\eqref{Hamilt} is integrable (it is the spin-1/2 XXZ chain in the spin language), and can be exactly solved using the Bethe ansatz~\cite{cazalilla_citro_review_11}. Next-nearest neighbor hoppings and interactions make the system nonintegrable. 

\section{Numerical Linked Cluster Expansions}\label{NLCE_sec}

NLCEs are based on the linked cluster theorem, which states that an extensive quantity per site $\mathcal{O}/N$, on a lattice with $N$ sites, can be calculated as the sum of the contributions from all connected clusters that can be embedded on the lattice:
\begin{equation}\label{nlce_eq}
\mathcal{O}/N=\sum_{c}\mathcal{L}(c)\times W_{\mathcal{O}}(c),
\end{equation}
where $\mathcal{L}(c)$ is the number of ways per site that cluster $c$ can be embedded on the lattice, and $W_{\mathcal{O}}(c)$ is the weight of observable $\mathcal{O}$ in cluster $c$. $W_{\mathcal{O}}(c)$ can be calculated, using the inclusion-exclusion principle, subtracting the weight of the observable in all the connected subclusters of $c$ to the value of the observable in cluster $c$:
\begin{equation}\label{weight_subtraction}
W_{\mathcal{O}}(c)=\mathcal{O}(c)- \sum_{s\subset c} W_{\mathcal{O}}(s).
\end{equation}
For the smallest cluster, $W_{\mathcal{O}}(c)=\mathcal{O}(c)$. 

The observable for a finite cluster with density matrix $\hat\rho_c$ is calculated as $\mathcal{O}(c)=\text{Tr}[\hat{O}\hat\rho_c]$. In high-temperature expansions, $\mathcal{O}(c)$ is calculated expanding $\hat \rho_c$ in powers of the inverse temperature $\beta=1/(k_BT)$. (We set $k_B=1$ in what follows.) In NLCEs, one calculates $\mathcal{O}(c)$ numerically using full exact diagonalization. 

When correlations are short ranged, the weights $W_{\mathcal{O}}(c)$ are expected to decrease rapidly with increasing the cluster size beyond the correlation length. Hence, calculating Eq.~\eqref{nlce_eq} to some finite order can be sufficient to estimate thermodynamic limit result to machine precision. One can reduce significantly the number of clusters to be diagonalized by identifying symmetries and topologies that relate clusters with identical expectation values of a given observable of interest. For a pedagogical introduction to NLCEs and their implementation, see Ref.~\cite{tang2013short}.

Another remarkable feature of NLCEs is that one has freedom to select different building blocks to construct the clusters to be used in the expansion, e.g., in the square lattice one can use bonds, sites, and squares  \cite{rigol2006numerical, *rigol2007numerical1, *rigol2007numerical2}. The order of the NLCE is then determined by the largest clusters having similar characteristics, e.g., in the square lattice it could be the number of bonds, sites, or squares, depending on the building blocks chosen. 

For the 1D lattice of interest in this work (see Fig \ref{1dlattice}), there are three straightforward ways of constructing the clusters. We will use two of them.

\begin{figure}[!t]
\centering
\includegraphics[width=80mm]{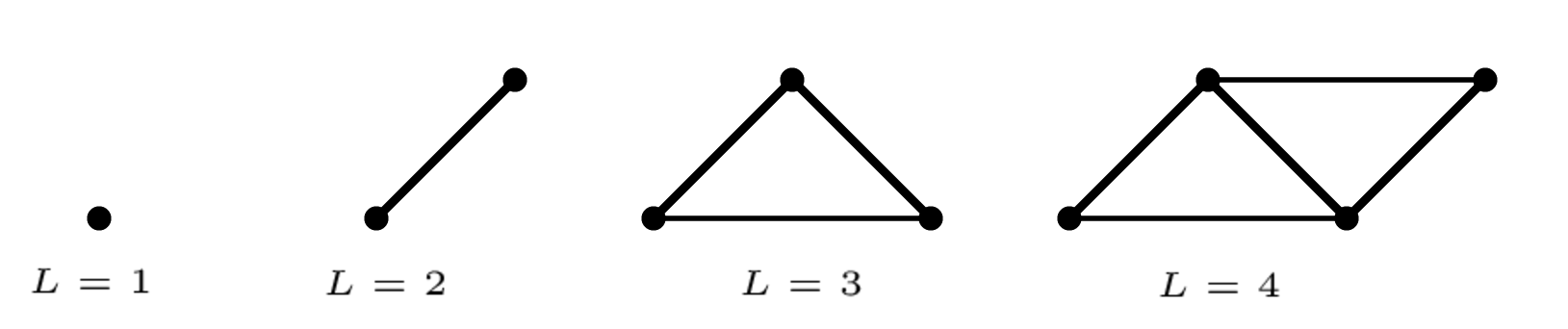}
\caption{The first four clusters of the maximally connected NLCE (NLCE-M).}\label{max_conn_fig}
\end{figure}

(i) \textbf{NLCE based on maximally connected clusters (NLCE-M)}: Starting from a single site, a nearest neighbor site is added each time along with all possible bonds it can have with the rest of the cluster (see Fig.~\ref{max_conn_fig}). This procedure generates clusters that have the maximum number of bonds for a given number of sites. The order of the ``maximally connected'' NLCE is set by the number of sites of the largest cluster (note that there is only one cluster for each given number of sites)~\cite{rigol2014quantum, rigol2016fundamental}. When only nearest neighbor interactions are present, this is the only NLCE possible for a 1D lattice~\cite{rigol2014quantum2, wouters_denardis_14_93, piroli_vernier_16}. Hence, this NLCE is expected to be best suited for weak next-nearest neighbor couplings.

\begin{figure}[h!]

\centering

\includegraphics[width=80mm]{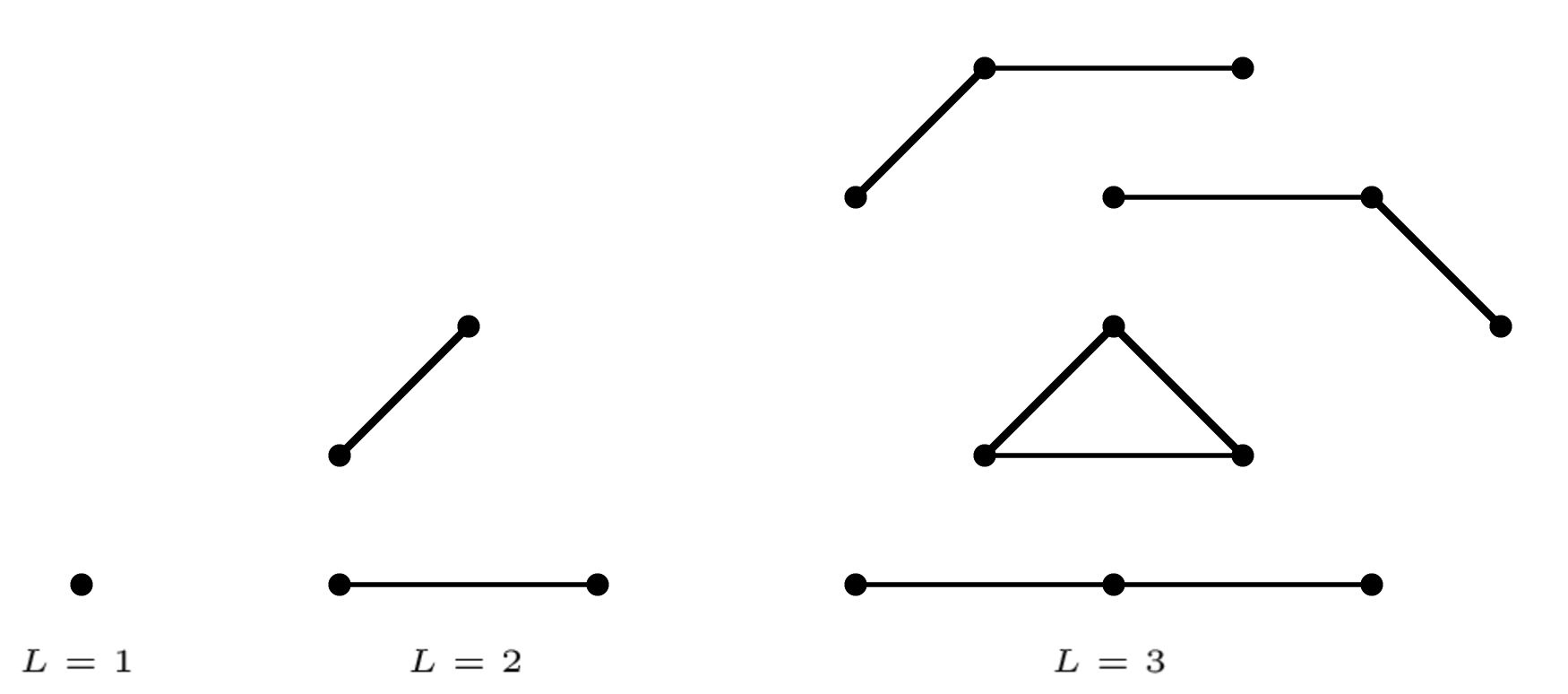}

\caption{Clusters with up to three sites in the site-based NLCE (NLCE-S).}\label{site_exp}

\end{figure}

(ii) \textbf{Site-based NLCE (NLCE-S)}: Starting from a single site, a nearest or next-nearest neighbor site is added one at a time along with all possible bonds it can have with the rest of the cluster (see Fig.~\ref{site_exp}). The order of the NLCE is set by the number of sites of the largest clusters (note that this time there are many clusters that can have the same number of sites). This NLCE is expected to outperform the maximally connected NLCE in the presence of strong next-nearest neighbor couplings. (This is apparent if one compares the clusters involved in both NLCEs for vanishing nearest neighbor couplings and nonvanishing next-nearest neighbor couplings.) However, this comes at a price as the number of clusters in the site-based NLCE grows exponentially with the order of the expansion. For a given number of sites $L$, there are $2^{L-1}$ clusters. Identifying which clusters are not related by symmetries (symmetry distinct clusters), allows one to reduce the number of clusters that need to be diagonalized by about a factor 2, as shown in Table~\ref{Table_ngraphs}. The top two clusters with three sites in Fig.~\ref{site_exp} are an example of clusters that are related by a symmetry (a reflection about the center of the next-nearest neighbor bond).

\begin{table}[!t]

\caption{Total number of clusters and of topological clusters in the site-based NLCE.}

\label{Table_ngraphs} 

\begin{ruledtabular}

\begin{tabular}{r r r}

	$L$	 &Symmetry distinct clusters	 & Total number of clusters \\

			\hline

	1           &1           &1      \\

	2           &2           &2      \\

	3           &3           &4      \\

	4           &6           &8      \\

	5          &10          &16      \\

	6          &20          &32      \\

	7          &36          &64      \\

	8          &72         &128      \\

	9         &136         &256      \\

	10        &272         &512      \\

	11        &528        &1024      \\

	12       &1056        &2048      \\

	13       &2080        &4096      \\

	14       &4160        &8192      \\

	15       &8256       &16384      \\

	16      &16512       &32768      \\

	17      &32896       &65536      \\

\end{tabular}

\end{ruledtabular}	

\end{table}

(iii) \textbf{Bond-based NLCE}: Starting from a single site, a bond with a nearest or next-nearest neighbor site is added one at a time. This NLCE includes all possible clusters that can be drawn on the lattice. The order of the NLCE is set by the number of bonds of the largest clusters, and the number of clusters that have a given number of bonds $N$ is $\sim 3^{N}$. The clusters in this NLCE are the ones that appear in the traditional high-temperature expansion. However, in the bond-based NLCE a full exact diagonalization calculation is carried out for each cluster, instead of the expansion in powers of $\beta$ carried out in the high-temperature expansion. A drawback of this expansion, when compared to the previous two, is that it is computationally more expensive as the number of clusters increases more rapidly with the number of bonds than for the other two with the number of sites. In addition, its convergence is generally worse than that of the site-based NLCE, as discussed in Ref.~\cite{rigol2006numerical, *rigol2007numerical1, *rigol2007numerical2} for various two-dimensional lattice geometries. Because of this, we do not consider the bond-based NLCE any further here.

\subsection{Resummations}\label{sec:resum}

To accelerate the convergence of NLCEs, we use Wynn's and Euler algorithms \cite{rigol2006numerical, *rigol2007numerical1, *rigol2007numerical2}.

First, we group together the contributions of all clusters with the same number of sites $L$
\begin{equation}\label{S_n}
 S_L=\sum_{\{c_L\}} \mathcal{L}(c_L)\times W_{\mathcal{O}}(c_L).
\end{equation}
Next, we make explicit that the sum in Eq.~\eqref{nlce_eq} can be carried out in clusters with up to $l$ sites, which gives the prediction of the $l^\text{th}$ order of the NLCE 
\begin{equation}
	\mathcal{O}_l/N= \sum_{L=1}^{l} S_L.
\end{equation}
The goal of resummation algorithms is to predict the result for $\mathcal{O}_{l\rightarrow\infty}$ from a finite sequence $\{\mathcal{O}_l\}$. 

Wynn's ($\epsilon$) algorithm is generally observed to give the best results for thermal states~\cite{rigol2006numerical, *rigol2007numerical1, *rigol2007numerical2}. In this resummation  algorithm, $\epsilon_n^{(k)}$ is defined as
\begin{eqnarray}
\epsilon_n^{(-1)}&=&0, \quad \epsilon_n^{(0)}=\mathcal{O}_n,\nonumber\\
\epsilon_n^{(k)}&=&\epsilon_{n+1}^{(k-2)}+\frac{1}{\epsilon_{n+1}^{(k-1)}-\epsilon_n^{(k-1)}}.
\end{eqnarray}
Here $k$ is the number of Wynn cycles. Only even cycles are expected to converge to the $l\rightarrow\infty$ result. Note that, every two cycles, the new sequence generated is reduced by two terms. We denote the estimate after $2k$ cycles as
\begin{equation}
\text{Wynn}_k(\mathcal{O}_l)=\epsilon_{l-2k}^{2k}.
\end{equation} 

We also find the Euler transformation to be useful, especially when studying quenches using the site-based NLCE. This resummation accelerates the convergence when the series $\{S_L\}$ alternates in sign. When the Euler resummation is used in the last $k$ terms of the series $\{S_L\}$, we write
\begin{eqnarray}
&&\text{Euler}_k(\mathcal{O}_l)=\sum_{L=0}^{l-k}S_L+(-1)^{l-k+1}\sum_{L=0}^{k}\frac{1}{2^{L+1}} T^l_{k,L},\\
&&\text{where\quad}
T^l_{k,L}=\left[\sum_{j=0}^{L}{L\choose j}S_{l-k+L-j+1}\right]. \nonumber
\end{eqnarray}

\subsection{Observables and convergence of NLCEs}\label{sec:observables}
We focus on a set of observables that includes: (1) The total energy per site $E=\text{Tr}[\hat{H}\hat\rho]/N$, (2) the variance of the energy per site $\Delta E^2=(\text{Tr}[\hat{H}^2\hat\rho]-(\text{Tr}[\hat{H}\hat\rho])^2)/N$, (3) the entropy per site $S=-\text{Tr}[\hat\rho\ln(\hat\rho)]/N$, (4) the total interaction energy per site $U=\text{Tr}[\hat{U}\hat\rho]/N$, where

\begin{eqnarray}
 \hat U&=&V\sum_{i}\left(\hat n_i-\frac{1}{2}\right)\left(\hat n_{i+1}-\frac{1}{2}\right) \nonumber\\
 &&+V'\sum_{i}\left(\hat n_i-\frac{1}{2}\right)\left(\hat n_{i+2}-\frac{1}{2}\right),
\end{eqnarray}
and (5) the occupation of the zero momentum mode, $m_{k=0}=\text{Tr}[\hat{m}_{k=0}\hat\rho]$. The momentum distribution $\hat m_{k}$ is the discrete Fourier transform of the one-body density matrix 
\begin{equation}
\hat m_{k}=\frac{1}{N}\sum_{j,j'} e^{ik(j-j')} \hat b_j^{\dagger}\hat b_{j'}^{}.
\end{equation}  
$m_{k}$ is of particular interest as it can be measured in experiments with ultracold atoms using time-of-flight expansion~\cite{bloch2008many}. Note that while $U$ is a local observable $m_k$ is a nonlocal one. In the expressions above, $\hat\rho$ stands for the many-body density matrix.

The convergence of the NLCE to the thermodynamic limit result is probed, for an observable $\mathcal{O}$, by calculating the normalized difference
\begin{equation}\label{rel_error_TH}
\Delta_l(\mathcal{O})=\left|\frac{\mathcal{O}_{l_\text{max}}-\mathcal{O}_{l}}{\mathcal{O}_{l_\text{max}}}\right|,
\end{equation}
where $l_\text{max}$ is the highest order of the NLCE that we are able to calculate. When $\Delta_l(\mathcal{O})$ reaches machine precision, one can think of the $l^\text{th}$ order of the NLCE as converged to the thermodynamic limit result. 

The site-based NLCE has a higher computational cost and a higher computational error than the maximally connected one due to the exponentially large number of clusters that need to be considered. Because of its computational cost, the site-based NLCE is carried out up to the $15^\text{th}$ order ($\text{NLCE-S}_{15}$), while the maximally connected one is carried out up to the $18^\text{th}$ order ($\text{NLCE-M}_{18}$). 

We also report results for exact diagonalization calculations in chains with periodic boundary conditions (ED-PBC). They are carried out in lattices with up to 20 sites ($\text{ED-PBC}_{20}$). In the ED-PBC calculations, we only consider even number of sites as those chains exhibit smaller finite-size effects. We compute the equivalent of Eq.~\eqref{rel_error_TH} as a function of the chain size $L$
\begin{equation}\label{rel_error_TH-ED}
\Delta_L(\mathcal{O})=\left|\frac{\mathcal{O}_{L_\text{max}}-\mathcal{O}_{L}}{\mathcal{O}_{L_\text{max}}}\right|.
\end{equation}

\section{Thermal equilibrium:\\ Grand canonical ensemble}\label{GE_sec}

In this section, we discuss the performance of the NLCEs for 1D systems in thermal equilibrium (in the grand canonical ensemble) described by Hamiltonian~\eqref{Hamilt}. We benchmark them against full exact diagonalization calculations in chains with periodic boundary conditions.

\begin{figure}[!t]
\includegraphics[width=.49\textwidth]{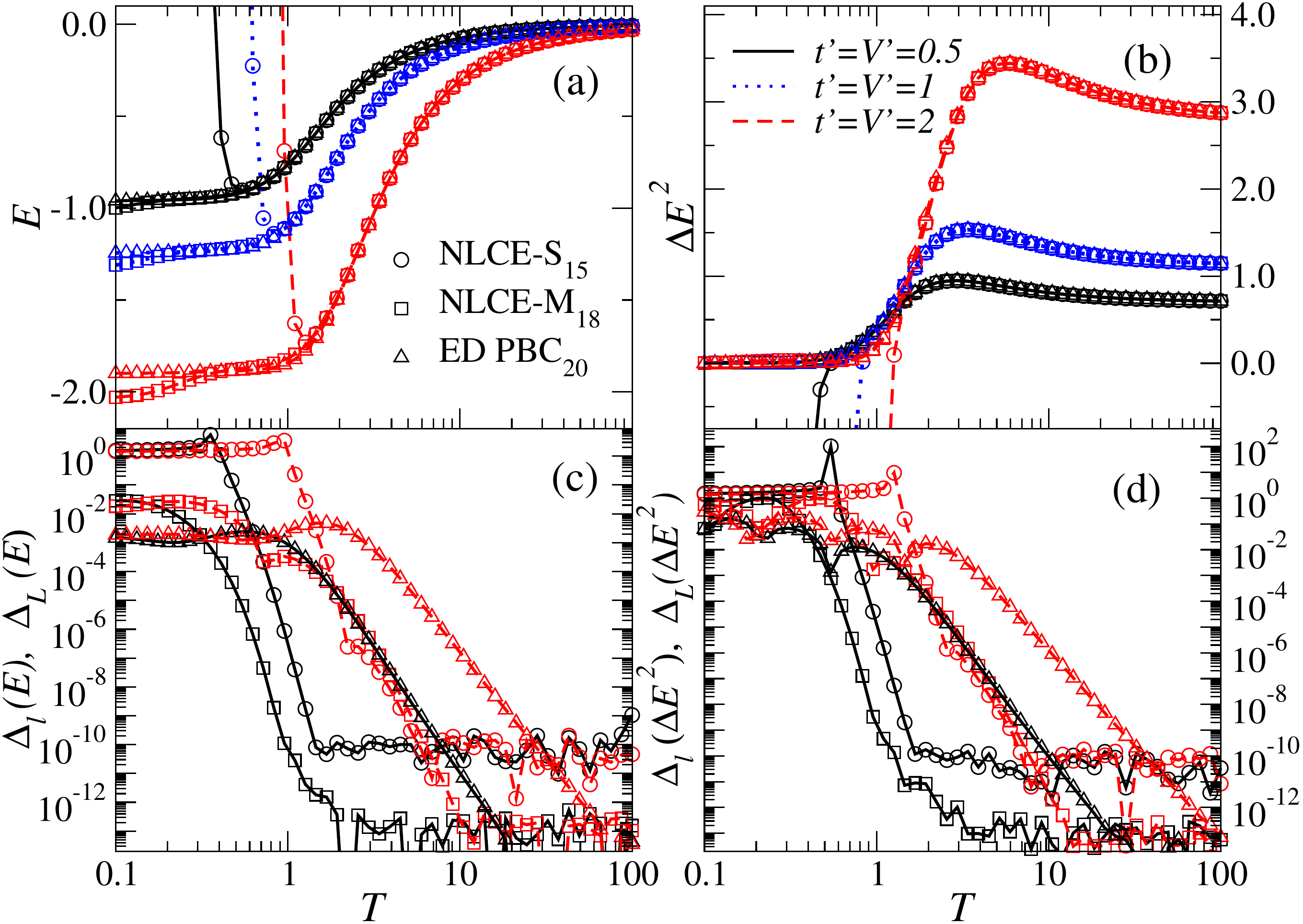}
\caption{(a) Energy $E$ and (b) variance of the energy $\Delta E^2$ vs $T$ for the last order of the NLCEs and for the largest chain for exact diagonalization. Normalized differences [Eq.~\eqref{rel_error_TH}] vs $T$ for: (c) the energy and (d) the variance of the energy. The normalized differences are between the last two orders of the NLCEs (17 and 18 for NLCE-M, and 14 and 15 for NLCE-S) and between the two largest chains (18 and 20 sites) for exact diagonalization. For the observables [(a) and (b)], results are reported for $t'=V'=0.5,$ 1, and 2. For the normalized differences [(c) and (d)], results are reported for $t'=V'=0.5$ and 2 (the ones for $t'=V'=1$ lie between those for $t'=V'=0.5$ and $2$ both for NLCEs and exact diagonalization).}\label{energy_T_GE}
\end{figure}

In the grand canonical ensemble, the density matrix of cluster $c$ is given by the expression

\begin{equation}\label{eq:GE}
 \hat\rho^\text{GE}_c=\frac{\exp(-\beta \hat H_c-\mu \hat N_c)}{\text{Tr}[\exp(-\beta \hat H_c-\mu \hat N_c)]},
\end{equation}
where $\hat{H}_c$ and $\hat{N_c}=\sum_{i\in c}\hat{n}_i$ are the Hamiltonian and the total particle number operator in cluster $c$, respectively. $\hat H_c$ is given by Eq.~\eqref{Hamilt} with all the nearest $(t,V)$ and next-nearest $(t',V')$ neighbor couplings present in the cluster. We set $t=V=1$, and focus on three sets of parameters $t'= V'=0.5,$ 1, and 2, which help understand the effect of increasing next-nearest neighbor couplings in the NLCE computations. We take $\mu=0$, which results in the systems being at half-filling as Hamiltonian~\eqref{Hamilt} is particle-hole symmetric.

\begin{figure}[!t]
\includegraphics[width=.49\textwidth]{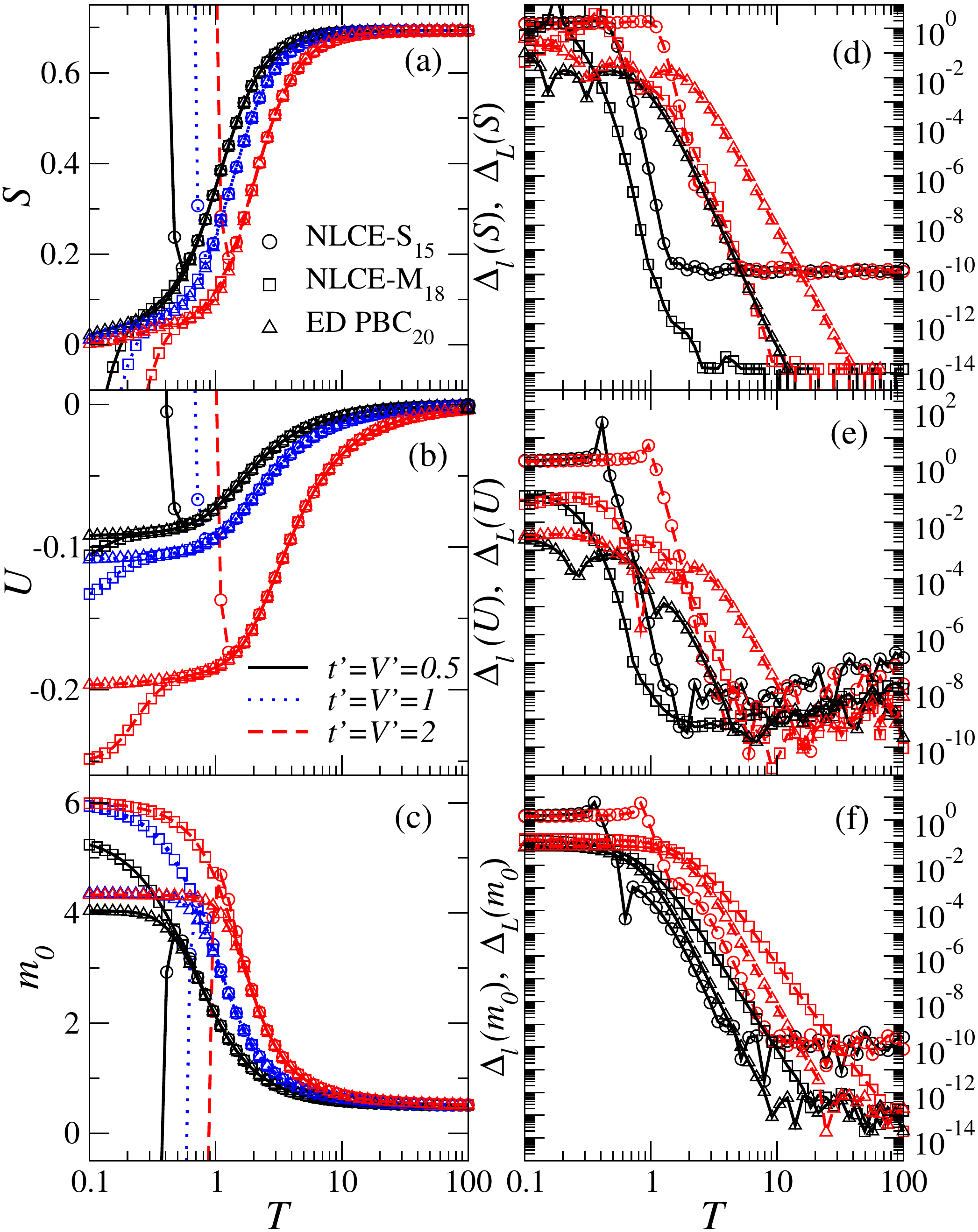}
\caption{Same as Fig.~\ref{energy_T_GE} for: (a) the entropy $S$, (b) the total interaction energy $U$, (c) the occupation of the zero momentum mode $m_{k=0}\equiv m_{0}$. The corresponding normalized differences are reported in panels (d)--(f).}\label{entropy_T_GE}
\end{figure}

In Figs.~\ref{energy_T_GE}(a) and \ref{energy_T_GE}(b), we plot the energy $E$ and the variance of the energy $\Delta E^2$ vs temperature as obtained using NLCEs and exact diagonalization for $t'=V'=0.5,$ 1, and 2. For each value of $t'=V'$, the plots are indistinguishable from each other for $T\gtrsim1$. Below $T\approx1$, the site-based NLCE exhibits a sharp divergence at a temperature that increases with increasing the value of $t'=V'$. The maximally connected NLCE and the exact diagonalization results differ from each other only for the energy $E$ at the lowest temperatures.

\begin{figure}[!t]

\includegraphics[width=.49\textwidth]{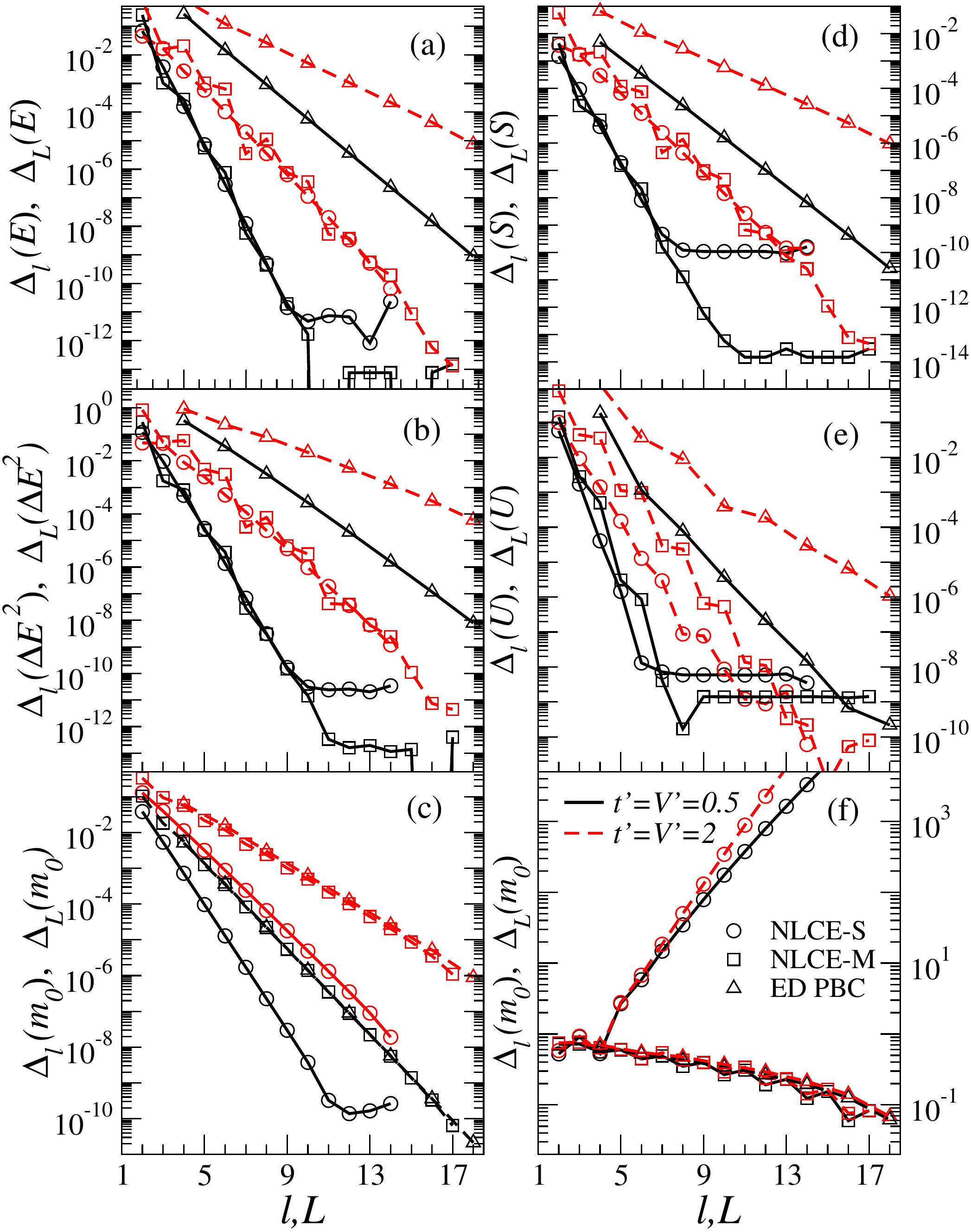}

\caption{Normalized differences $\Delta_l(\mathcal{O})$ vs $l$ for the maximally connected and site-based NLCEs, and $\Delta_L(\mathcal{O})$ vs $L$ for the exact diagonalization calculations. In each panel, results are reported for $t'=V'=0.5$ and 2 at fixed temperature. (a)--(e) Results for $T=5.96$, a temperature at which both NLCEs converge to machine precision. (f) Results for $T=0.1$, a temperature at which the site-based NLCE diverges [$\Delta_l(m_0)$ in the site-based NLCE was calculated using the result for the $18^\text{th}$ order of the maximally connected NLCE].}\label{energy_ns_GE}

\end{figure}

The normalized differences [Eq.~\eqref{rel_error_TH}] for $E$ and $\Delta E^2$ vs temperature, between the last two orders of the NLCEs and between the two largest chains in the exact diagonalization, are reported in Figs.~\ref{energy_T_GE}(c) and \ref{energy_T_GE}(d) for $t'=V'=0.5$ and 2. For each value of $t'=V'$, one can see that the exact diagonalization results are the first to depart from machine precision as the temperature is decreased. The maximally connected and site-based NLCEs results depart from machine precision about the same temperature for $t'=V'=2$, while for $t'=V'=0.5$ the site-based NLCE departs from machine precision at higher temperature than the maximally connected one. The results for $t'=V'=1$ (not shown) are in between.

\begin{figure}[t]

\centering

\includegraphics[width=.44\textwidth]{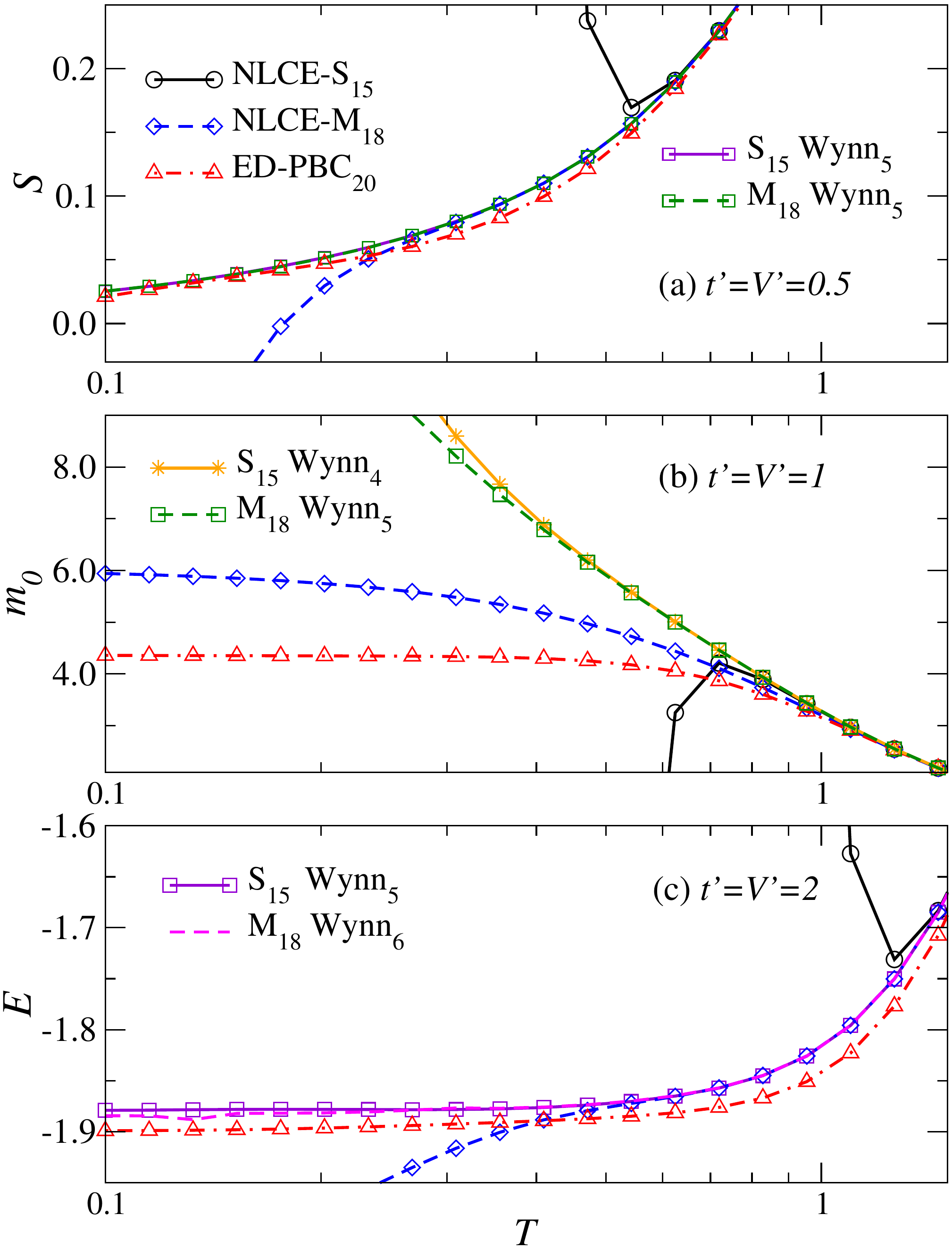}

\caption{Wynn$_k$ resummations for the site-based (NLCE-S$_{15}$) and maximally connected (NLCE-M$_{18}$) NLCEs, along with the results for the last order of the NLCEs and the largest chains in the exact diagonalization calculations. (a) Entropy for $t'=V'=0.5$, (b) Occupation of the zero momentum mode for $t'=V'=1$, and (c) Energy for $t'=V'=2$.}\label{entropy_extrp_T_GE}

\end{figure}

The results obtained for other observables are qualitatively similar to the ones for the energy and the variance of the energy. In Fig.~\ref{entropy_T_GE}, we report results for the entropy $S$, the total interaction energy $U$, and the occupation of the zero momentum mode $m_{0}$. The main difference between those results and the ones for the energy and the variance of the energy is seen for $m_{0}$. Figure~\ref{entropy_T_GE}(f) shows that the normalized differences of the site-based NLCE are the ones that, when increasing $T$, reach machine precision at the lowest temperature. Namely, the site-based NLCE is the one that converges at the lowest temperatures.

Next one can ask, for a given temperature at which the NLCEs converge to machine precision, how the converged result is obtained with increasing the order of the expansion. In Fig.~\ref{energy_ns_GE}(a)--\ref{energy_ns_GE}(e), we plot the normalized distances for the five observables studied in Figs.~\ref{energy_T_GE} and \ref{entropy_T_GE} as a function of the order $l$ of the NLCEs and of the chain size $L$ of the exact diagonalization. We report results for $t'=V'=0.5$ and 2, and for $T=5.96$, for which the NLCEs have converged to machine precision for all observables studied. Figures~\ref{energy_ns_GE}(a)--\ref{energy_ns_GE}(e) show that: (1) the convergence of NLCEs (exact diagonalization) with increasing the order of the expansion (the chain size) is exponential~\cite{iyer2015optimization}, (2) for $E$, $\Delta E^2$, and $S$, the convergence of the maximally connected and the site-based NLCEs is similar, and they converge more rapidly than the exact diagonalization calculations, (3) for $U$ and $m_0$, the site-based NLCE exhibits the fastest convergence, followed by the maximally connected NLCE and exact diagonalization calculations, and (4) in all calculations, increasing the strength of the next-nearest neighbor couplings slows down the convergence.

Qualitative differences between the site-based NLCE, the maximally connected NLCE, and exact diagonalization mostly occur at low temperatures, at which the site-based NLCE diverges. In Fig.~\ref{energy_ns_GE}(f), we plot the normalized differences for the zero momentum mode occupation at $T=0.1$. [$\Delta_l(m_0)$ in the site-based NLCE was computed using the result for the $18^\text{th}$ order of the maximally connected NLCE.] The normalized differences decrease for the maximally connected NLCE (with $l$) and the exact diagonalization (with $L$) calculations, but increase nearly exponentially with $l$ for the site-based NLCE. We discuss the origin of this divergence in Sec.~\ref{weight_sec}.

Whenever NLCE calculations fail to converge, one can use resummations to accelerate the convergence. In Fig.~\ref{entropy_extrp_T_GE}(a)--\ref{entropy_extrp_T_GE}(c), we show $S$, $m_0$, and $E$ for $t'=V'=0.5$, 1 and 2, respectively, after various Wynn resummation cycles (see Sec.~\ref{sec:resum}) applied to both the site-based and the maximally connected NLCEs. Wynn's algorithm accelerates the convergence the most for systems in thermal equilibrium. It is remarkable that the resummed results of the two NLCEs agree with each other for all observables and most temperatures reported in Fig.~\ref{entropy_extrp_T_GE}. They also agree with the maximally connected NLCE results below temperatures at which the site-based NLCE exhibits a divergence, and are clearly different from the exact diagonalization ones, which suffer from finite-size effects. Having an agreement between resummed results of two different NLCEs suggests that the resummed results are converged and that resummations allow one to extend the regime of applicability of the NLCEs beyond the one provided by the bare NLCE sums in Eq.~\eqref{nlce_eq}.

\section{Quantum Quenches}\label{DE_sec}

In this section, we study quantum quenches starting from initial thermal equilibrium states. Namely, the initial state is in thermal equilibrium for an initial Hamiltonian $\hat H^{I}$, at temperature $T_{I}$ [$\beta_{I}=(T_{I})^{-1}$] and chemical potential $\mu_{I}$. This means that, in each cluster $c$ of the NLCEs, the initial density matrix in the grand canonical ensemble is
\begin{equation}
\hat \rho^{I}_c=\frac{\exp(-\beta_{I} \hat H^{I}_c-\mu_{I} \hat N_c)}{\text{Tr}[\exp(-\beta_{I} \hat H^{I}_c-\mu_{I} \hat N_c)] }.
\end{equation}
As in Sec.~\ref{GE_sec}, $\hat{H}^{I}_c$ and $\hat{N}_c$ stand for the Hamiltonian and total particle number operator in cluster $c$, respectively.

The quench consists of changing the initial Hamiltonian $\hat{H}^I$ ($\hat{H}^I_c$ for cluster $c$) into a new time-independent (final) Hamiltonian $\hat H$ ($\hat{H}_c$ for cluster $c$) and at the same time disconnecting the system from the bath. As a result, the dynamics that follows the quench is unitary, and,  as a function of time $\tau$, the density matrix of cluster $c$ can be written as (we set $\hbar=1$)
\begin{eqnarray}
\hat \rho_c(\tau)&=&e^{-i\hat H_c \tau}\hat\rho^I_c e^{iH_c\tau}\nonumber\\
 &=& \sum_{m,n}e^{-i(E_m-E_n)\tau}\ket{m}\bra{m}\hat\rho^I_c\ket{n}\bra{n},\label{rho_time}
\end{eqnarray}
where $\ket{m}$ and $E_m$ are the energy eigenkets and eigenvalues for $\hat H_c$, respectively. The expectation value of an observable at time $\tau$ in cluster $c$ is given by $\mathcal{O}_c(\tau)=\text{Tr}[\hat\rho_c(\tau)\hat{O}]$. Here, we are interested in the infinite-time average $\bar{\mathcal{O}}_c=\lim_{\tau'\rightarrow\infty}(1/\tau')\int_{0}^{\tau'}O_c(\tau)d\tau$, which describes observables after relaxation \cite{dalessio_kafri_16}. One can also write $\bar{\mathcal{O}}_c= \text{Tr}[\hat\rho^\text{DE}_c\hat{O}]$, where $\hat\rho^\text{DE}_c$ is the diagonal ensemble \cite{rigol2008thermalization} density matrix in cluster $c$
\begin{equation}
\hat \rho^\text{DE}_c=\lim_{\tau'\rightarrow\infty}\frac1{\tau'}\int_{0}^{\tau'}\hat\rho_c(\tau) d\tau=\sum_{m}(\bra{m}\hat \rho^I_c\ket{m})\ket{m}\bra{m},\label{DE equation}
\end{equation}
assuming that there are no degeneracies in the energy spectrum. This is the density matrix used in the NLCEs to obtain observables after relaxation in the thermodynamic limit following a quantum quench. The entropy in the diagonal ensemble \cite{rezek2006irreversible,polkovnikov_11, santos_polkovnikov_11}, also known as the diagonal entropy, is computed as $S^\text{DE}_c=-\text{Tr}[\hat\rho^\text{DE}_c\ln\hat \rho^\text{DE}_c]$.

For nonintegrable systems, which are the ones of interest in this work, it has been shown that the NLCE results for the diagonal ensemble agree with those of the grand canonical ensemble with the temperature $T$ and chemical potential $\mu$ [see Eq.~\eqref{eq:GE}] selected such that \cite{rigol2014quantum, rigol2016fundamental}
\begin{eqnarray}
 \text{Tr}[\hat\rho^\text{GE}\hat H]&=&\text{Tr}[\hat \rho^{I}\hat H]\label{finalT},\\
 \text{Tr}[\hat\rho^\text{GE}\hat N]  &=&\text{Tr}[\hat \rho^{I}\hat N],
\end{eqnarray}
which means that the observables studied thermalize. 

In what follows, we restrict our analysis to half-filled systems, which is achieved by setting $\mu_{I}=\mu=0$ (the total number of particles per site is conserved during the quench). Also, the initial Hamiltonian $\hat H^{I}$ is always taken to have only nearest neighbor couplings $t_I=0.5$ and $V_I=1.5$ ($t'_I=V'_I=0$). After the quench, the Hamiltonian parameters change to $t=V=1$, and $t'=V'=0.5$, 1, and 2 (the same parameters considered for systems in thermal equilibrium in Sec.~\ref{GE_sec}). 

\begin{figure}[!t]
\includegraphics[width=.47\textwidth]{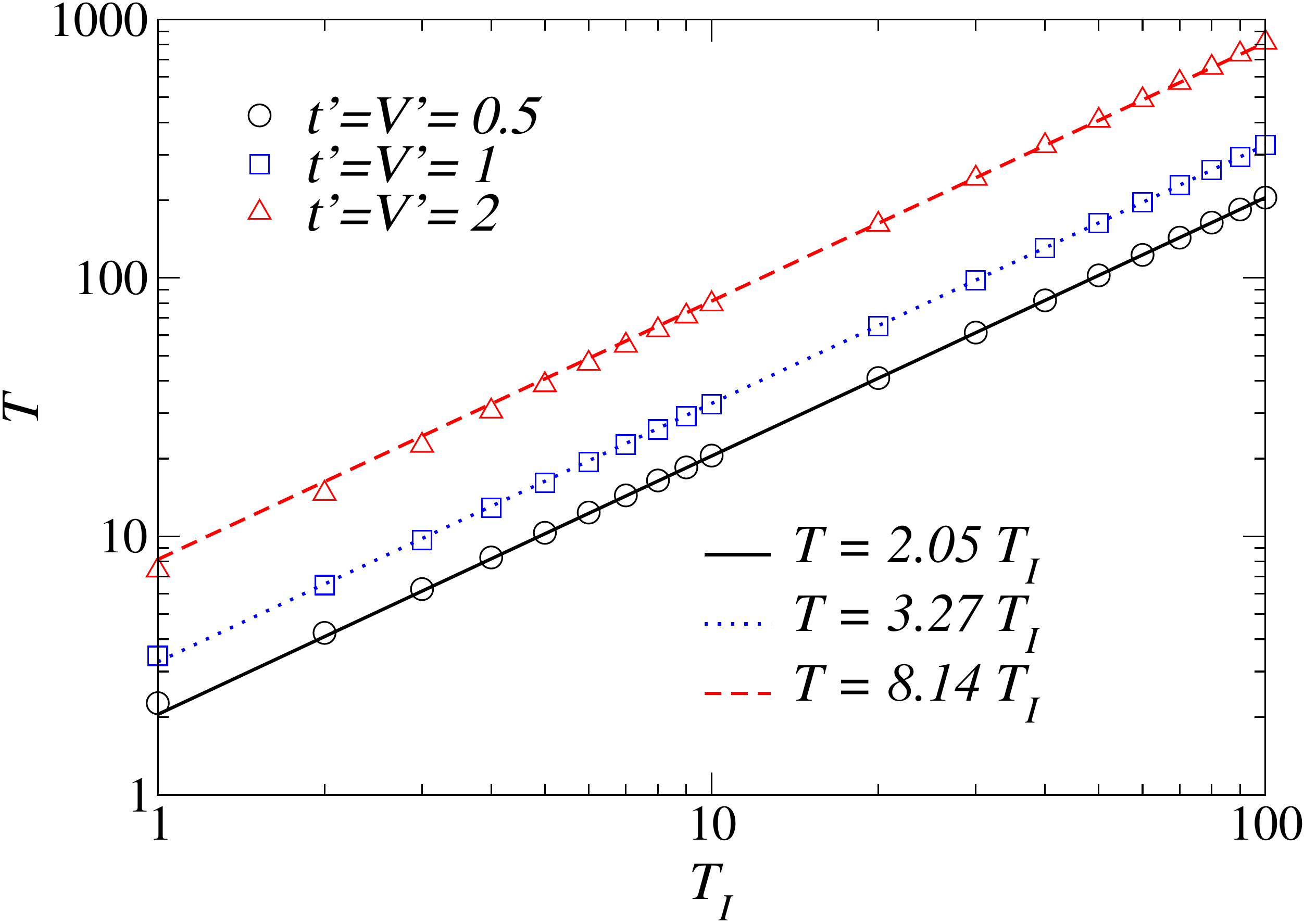}
\caption{Final temperature $T$ as a function of the initial temperature $T_I$ for quenches with $t_I=0.5$, $V_I=1.5$, $t_I'=V_I'=0$, and $t=V=1$, $t'=V'=0.5,$ 1, and 2. Also shown is a linear fit  to the data points with $T_I\geq 10$.}\label{Ti_Tf}
\end{figure}

The temperature of the grand canonical ensemble that describes observables after relaxation following the quench (in short, the ``final effective temperature'') $T$ is calculated using the maximally connected NLCE requiring that the relative difference between the total energy per site predicted within the diagonal and grand canonical ensembles for $l=18$ is below $10^{-12}$. In Fig.~\ref{Ti_Tf}, we plot $T$ vs $T_I$ for the quenches of interest in this work. For large values of $T_I$, the dependence of $T$ on $T_I$ is an almost linear one (see the fits in Fig.~\ref{Ti_Tf}).

\begin{figure*}[!t]
\includegraphics[width=.8\textwidth]{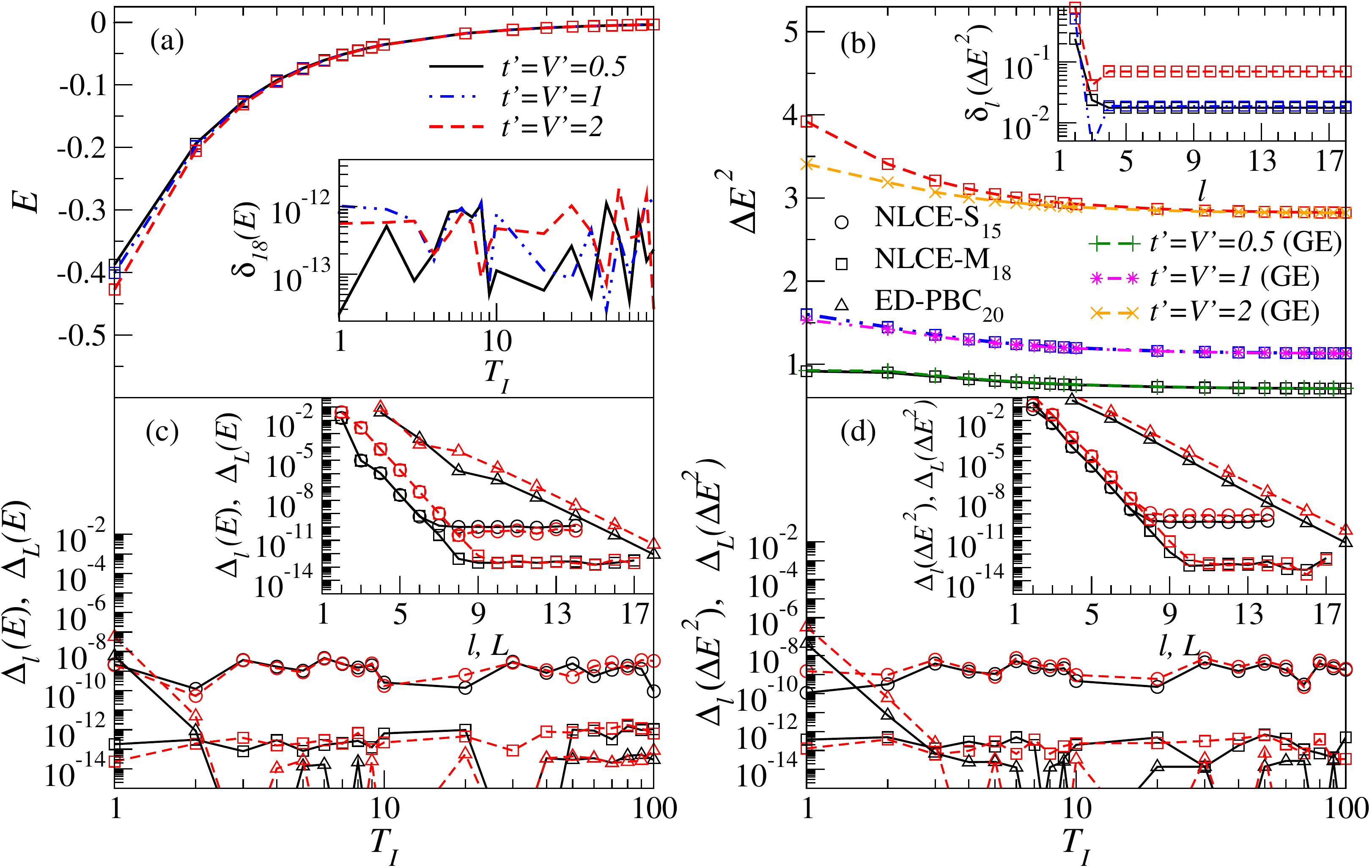}
\caption{(a) $E$ in the diagonal ensemble, and (b) $\Delta E^2$ in the diagonal and grand canonical ensembles, calculated using the $18^\text{th}$ order of the maximally connected NLCE. Inset in (a), $\delta_{18}(E)$ vs $T_I$. Inset in (b), $\delta_{l}(\Delta E^2)$ vs $l$ for $T_I=2$. See Eq.~\eqref{rel_error_DH} for the definition of $\delta_l(\mathcal{O})$. (c) $\Delta_{l=14}(E)$ for NLCE-S, $\Delta_{l=17}(E)$ for NLCE-M, $\Delta_{L=18}(E)$ for ED-PBC vs $T_I$ and (d) $\Delta_{l=14}(\Delta E^2)$ for NLCE-S, $\Delta_{l=17}(\Delta E^2)$ for NLCE-M, $\Delta_{L=18}(\Delta E^2)$ for ED-PBC vs $T_I$, while their insets show results for $\Delta_{l}(E)$, $\Delta_{L}(E)$ and $\Delta_{l}(\Delta E^2)$, $\Delta_{L}(\Delta E^2)$, respectively, vs $l$ and $L$ for $T_I=2$. See Eqs.~\eqref{rel_error_TH} and \eqref{rel_error_TH-ED} for the definition of $\Delta_{l}({\mathcal O})$ and $\Delta_{L}({\mathcal O})$. In all quenches, $t_I=0.5, V_I=1.5$, and $t_I'=V_I'=0$. In (a) and (b), we show results for $t'=V'=0.5$, 1, and 2, while in (c) and (d), we show results for $t'=V'=0.5$ and 2.}\label{energy_T_ns_DE}
\end{figure*}

Quantities such as the energy $E$, the variance of the energy $\Delta E^2$, and the variance of the total particle number $\Delta N^2$ are conserved during the quench, e.g., $\text{Tr}[\hat\rho^{I}\hat H]=\text{Tr}[\hat \rho^\text{DE}\hat H]$ (note that this is different from $\text{Tr}[\hat\rho^{I}\hat H^I]$). This means that they converge exponentially fast with the order of the NLCEs, as discussed in Sec.~\ref{GE_sec} for various observables in systems in thermal equilibrium and as seen in Refs.~\cite{rigol2014quantum, rigol2016fundamental} for quantum quenches. Also, since there are no next-nearest neighbor couplings in $\hat H_I$, the site-based and maximally connected NLCEs provide identical results within machine precision (the clusters with nonzero weight in both expansions are the same).

In Fig.~\ref{energy_T_ns_DE}(a) and \ref{energy_T_ns_DE}(b), we show $E$ and $\Delta E^2$ calculated using the maximally connected NLCE in the diagonal ensemble vs $T_I$. The energies after the quench are rather close to each other for the values of $t'=V'$ considered here, but the variance of the energy increases significantly as the value of $t'=V'$ is increased. Results for the normalized differences between the last two orders of the maximally connected and site-based NLCEs are presented in the main panels of Fig.~\ref{energy_T_ns_DE}(c) and \ref{energy_T_ns_DE}(d) for $E$ and $\Delta E^2$, respectively. They can be seen to be within machine precision for the temperatures considered. The insets in Fig.~\ref{energy_T_ns_DE}(c) and \ref{energy_T_ns_DE}(d) also show normalized differences but plotted as a function of the order of the NLCE for a fixed temperature $T_I=2$. As expected, the site-based and the maximally connected NLCEs give identical results at low orders. However, as the order is increased, the site-based NLCE saturates at a higher normalized difference (machine precision for this expansion) because it involves exponentially many more clusters than the maximally connected one.

To determine whether observables in the diagonal ensemble approach the grand canonical ensemble predictions as the order of the NLCE is increased, we compute the normalized difference
\begin{equation}\label{rel_error_DH}
\delta_l(\mathcal{O})=\left|\frac{\mathcal{O}_{18}^\text{GE}-\mathcal{O}_{l}^\text{DE}}{\mathcal{O}_{18}^\text{GE}}\right|
\end{equation}
The value of this quantity for the energy must be within machine precision in the $18^\text{th}$ order of the maximally connected NLCE, as the temperature is determined so that $\delta_{l=18}(E)<10^{-12}$. The inset in Fig.~\ref{energy_T_ns_DE}(a) shows that this is indeed the case in our calculations. On the other hand, the variance of the energy after the quench does not generally agree with that of the corresponding thermal equilibrium ensemble. This can be seen in the main panel in Fig.~\ref{energy_T_ns_DE}(b), in which we also plot the grand canonical ensemble predictions for $\Delta E^2$, and in the inset in Fig.~\ref{energy_T_ns_DE}(b), in which we plot $\delta_l(\Delta E^2)$ vs $l$ for $T_I=2$. This is the result of $\Delta E^2$ being a conserved quantity during the quench. As discussed in Ref.~\cite{rigol2014quantum}, $\Delta E^2$ can be used to distinguish the diagonal ensemble from thermal ensembles. We note that an agreement between $\Delta E^2$ in the diagonal and thermal ensembles is not required for few-body observables to thermalize. So long as the variance of the energy is extensive (as it is in all the quenches considered here), such that its square root (the width of the energy distribution) is subextensive, and eigenstate thermalization occurs for a given observable, then the observable is guarantied to thermalize in the thermodynamic limit~\cite{dalessio_kafri_16}.

\begin{figure*}
\includegraphics[width=.915\textwidth]{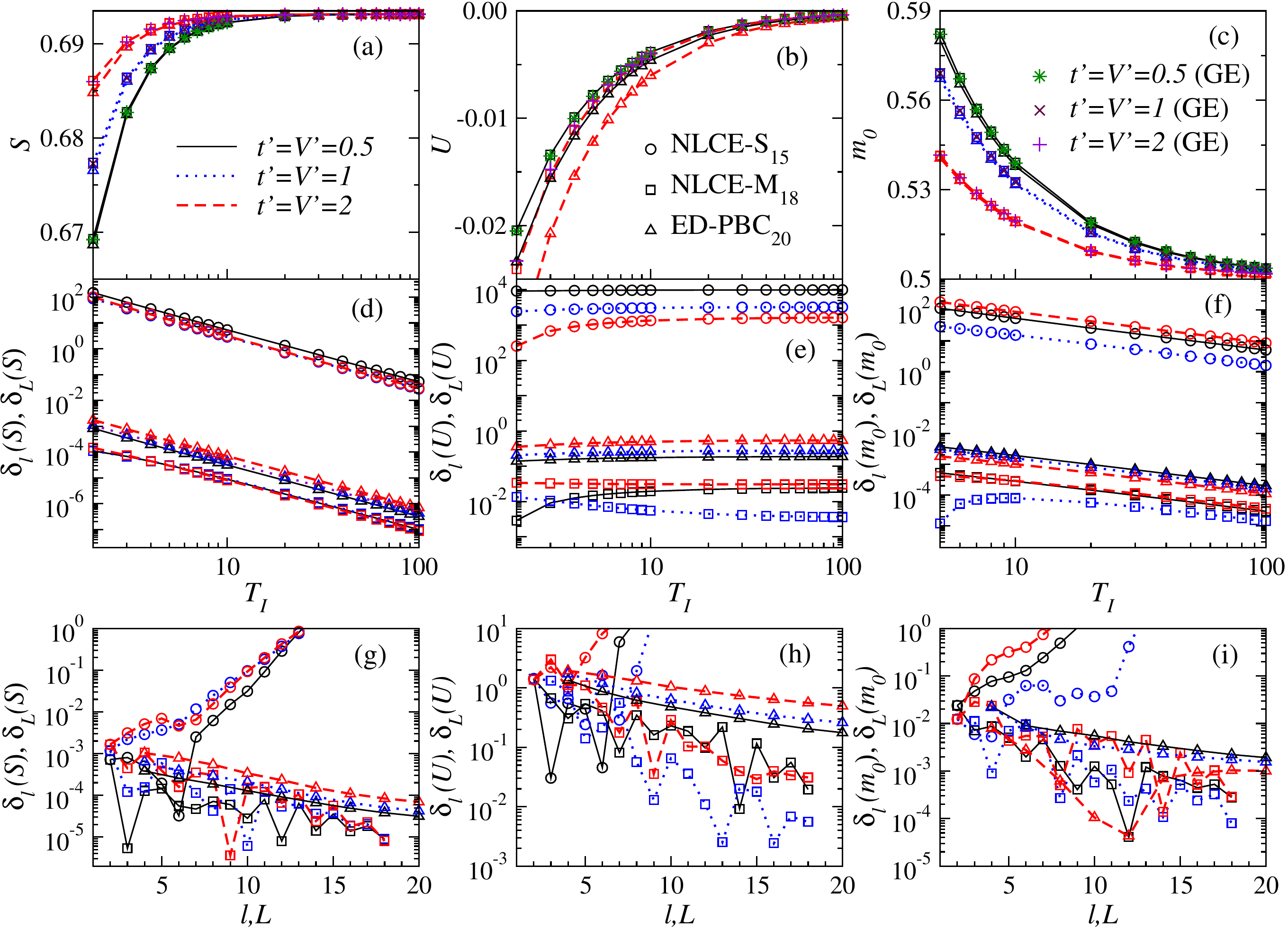}
\caption{(a)--(c) Entropy, interaction energy, occupation of the zero momentum mode in the diagonal ensemble, respectively, calculated using the maximally connected NLCE (NLCE-M$_{18}$) and exact diagonalization (ED-PBC$_{20}$). We also show results in the corresponding grand canonical ensemble obtained using the maximally connected NLCE (NLCE-M$_{18}$). (d)--(f) Normalized differences, as defined in Eq.~\eqref{rel_error_DH} and Eq.~\eqref{rel_error_DH-ED}, for the same observables as in (a)--(c), obtained using the maximally connected and site-based NLCEs as well as exact diagonalization. Results are reported for the highest order of the NLCEs and the largest chain for exact diagonalization. (g)--(i): Normalized differences vs the order $l$ of the NLCEs and the chain size $L$ in the exact diagonalization, at fixed initial temperature $T_I=10$. In all panels but (b), we show results for quenches with $t'=V'=0.5$, 1, and 2. For clarity, in (b) we only show results for $t'=V'=0.5$ and 2.}\label{entropy_T_ns_DE}
\end{figure*}

For observables that are not conserved during the quench, the overwhelming majority of observables of interest, the convergence of NLCEs is slower for the diagonal ensemble than for the grand canonical one \cite{rigol2014quantum}. In what follows, we focus on the behavior of the entropy $S$ (a thermodynamic observable), the potential energy $U$ (a local observable), and the occupation of the zero momentum mode $m_0$ (a nonlocal observable). Since we restrict our analysis to quenches in which the final Hamiltonian has $t'=V'\neq0$, namely, a nonintegrable Hamiltonian, those observables are expected to thermalize \cite{dalessio_kafri_16}. Hence, we can use converged results from the corresponding grand canonical ensemble to study the convergence of the NLCEs for the diagonal ensemble. This means that the normalized difference that will be the focus of the analysis that follows is the one given by Eq.~\eqref{rel_error_DH}. In all calculations of $\delta_l(\mathcal{O})$ in the maximally connected and site-based NLCEs, as well as of 
\begin{equation}\label{rel_error_DH-ED}
\delta_L(\mathcal{O})=\left|\frac{\mathcal{O}_{18}^\text{GE}-\mathcal{O}_{L}^\text{DE}}{\mathcal{O}_{18}^\text{GE}}\right|,
\end{equation}
in exact diagonalization, we use $\mathcal{O}_{18}^\text{GE}$ obtained from the maximally connected NLCE. We restrict our analysis to initial temperatures such that, for the grand canonical ensemble that describes the systems after the quench, $\Delta_{17}( S) \lesssim 10^{-10}$, $\Delta_{17}(m_0) \lesssim 10^{-10}$, and $\Delta_{17}(U)\lesssim 10^{-8}$.

In Figs.~\ref{entropy_T_ns_DE}(a)--\ref{entropy_T_ns_DE}(c), we show $S$, $U$ and $m_0$ in the diagonal ensemble as obtained using the  $18^\text{th}$ order of the maximally connected NLCE and exact diagonalization in a periodic chain with 20 sites (the results of the site-based NLCE for the diagonal ensemble fall beyond the scale of these plots). We also report the results obtained for the same observables in the grand canonical ensemble using the $18^\text{th}$ order of the maximally connected NLCE. The agreement between the diagonal and grand canonical ensemble results within the maximally connected NLCE is excellent. Differences between those two and exact diagonalization results are only apparent for $S$ and $m_0$ at the lowest temperatures, and for $U$ at most temperatures.

In Figs.~\ref{entropy_T_ns_DE}(d)--\ref{entropy_T_ns_DE}(f), we show the normalized differences $\delta_l(S)$, $\delta_l(U)$ and $\delta_l(m_0)$ for the last order of the NLCE calculations and $\delta_L(S)$, $\delta_L(U)$ and $\delta_L(m_0)$ for the largest chain in the exact diagonalization calculations. Those differences can be seen to decrease with increasing $T_I$ for $S$ and $m_0$, as expected as increasing $T_I$ should decrease the convergence (NLCE) and finite-size (exact diagonalization) errors. The differences remain fairly constant for $U$ because $U_{18}^\text{GE}$ [which is in the denominator of $\delta_l(U)$] approaches zero with increasing $T_I$. The other fact that is apparent in Fig.~\ref{entropy_T_ns_DE}(d)--\ref{entropy_T_ns_DE}(f) is that the maximally connected NLCE exhibits normalized differences that are between one and two orders of magnitude smaller than those of the exact diagonalization, and that the site-based NLCE exhibits differences that are several orders of magnitude larger at all initial temperatures. The latter is qualitatively different from what was discussed for systems in thermal equilibrium in the context of Fig.~\ref{entropy_T_GE}. 

Figures~\ref{entropy_T_ns_DE}(g)--\ref{entropy_T_ns_DE}(i) depict the same normalized differences but as a function of $l$ (NLCEs) and $L$ (exact diagonalization), at a fixed initial temperature $T_I=10$. Those plots make apparent that while the maximally connected NLCE and exact diagonalization calculations converge (though more slowly than for systems in thermal equilibrium) with increasing the order of the expansion and the chain sizes, respectively, the site-based NLCE diverges even at this relatively high initial temperature. We discuss the origin of this divergence in the next section.

\begin{figure}
\includegraphics[width=.45\textwidth]{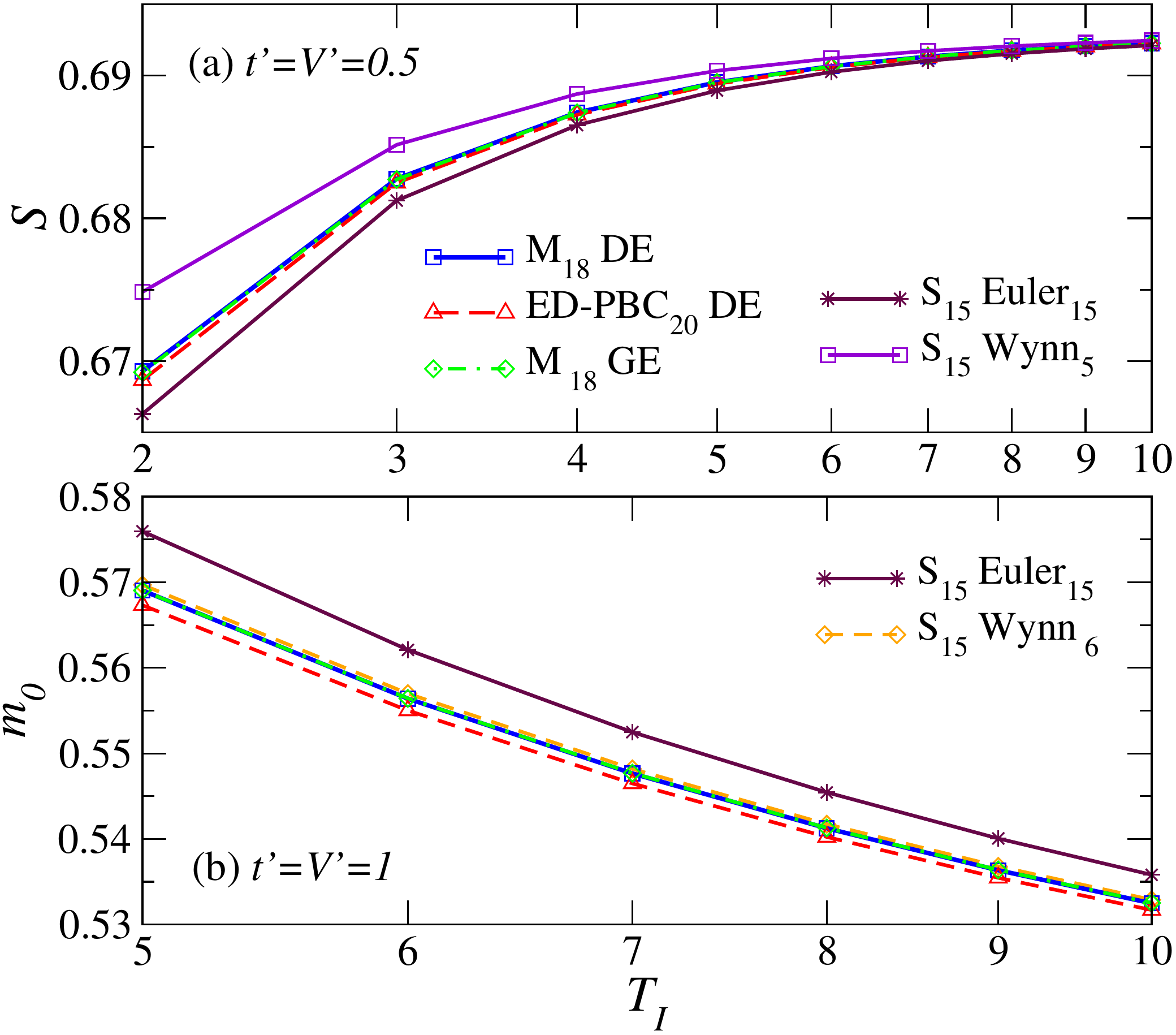}
\caption{Wynn and Euler resummations of the site-based NLCE for the diagonal ensemble, together with the $18^\text{th}$ order of the maximally connected NLCE for the diagonal and grand canonical ensembles, and with exact diagonalization results for the diagonal ensemble in a chain with 20 sites. (a) Entropy for $t'=V'=0.5$, and (b) occupation of the zero momentum mode for $t'=V'=1$.}\label{DE_T_extrp}
\end{figure}

As for systems in thermal equilibrium, one can use resummations to accelerate the convergence of NLCEs for the diagonal ensemble. In Fig.~\ref{DE_T_extrp}, we show results for resummations of the site-based NLCE for the entropy [Fig.~\ref{DE_T_extrp}(a)] and the occupation of the zero momentum mode [Fig.~\ref{DE_T_extrp}(b)]. The results after resummation can be seen to be close to the thermal equilibrium predictions, and in some instances [such as the $6^\text{th}$ cycle of the Wynn algorithm for $m_0$ in Fig.~\ref{DE_T_extrp}(b)] they are actually closer to the thermal equilibrium predictions than the exact diagonalization ones. However, the direct sums for the maximally connected NLCE exhibit a better agreement with the thermal equilibrium predictions than the resummed results for the site-based NLCE. In summary, our study indicates that the maximally connected NLCE is the best approach (among the ones considered in this work) for diagonal ensemble calculations after quantum quenches in the thermodynamic limit.

\section{Divergence of the site-based NLCE}\label{weight_sec}

In the systems in thermal equilibrium discussed in Sec.~\ref{GE_sec}, we have seen that below some temperature the site-based NLCE diverges, while for quantum quenches it appears to diverge for all initial temperatures considered. Here, we provide an understanding of the origin of those divergences, and their differences, by studying the average weight of the clusters with a given number of sites, and comparing it to the number of clusters with that number of sites. 

The average cluster weight $W(\mathcal{O},L)$ [see Eq.~\eqref{nlce_eq}], of an observable $\mathcal{O}$ in clusters with $L$ sites, is
\begin{equation}
W(\mathcal{O},L)=\frac{\sum_{\{c_L\}}{\mathcal L}(c_L)\times W_{\mathcal{O}}(c_L)}{\sum_{\{c_L\}}{\mathcal L}(c_L)}.
\end{equation} 
When the clusters in the NLCE are not large enough compared to the correlation length associated with a given observable, the weights of the observable in those clusters need not be small even if they decrease with increasing the cluster size. Since the number of clusters in the site-based NLCE grows exponentially fast with the number of sites in the clusters, the combination of non-rapidly-enough decreasing weights with the exponential increase of the number of clusters can lead to a divergence. For the maximally connected NLCE, there is only one cluster with any given number of sites. Hence, no divergence is expected with increasing cluster size no matter the cluster sizes considered. 

\begin{figure}[!b]
\includegraphics[width=.49\textwidth]{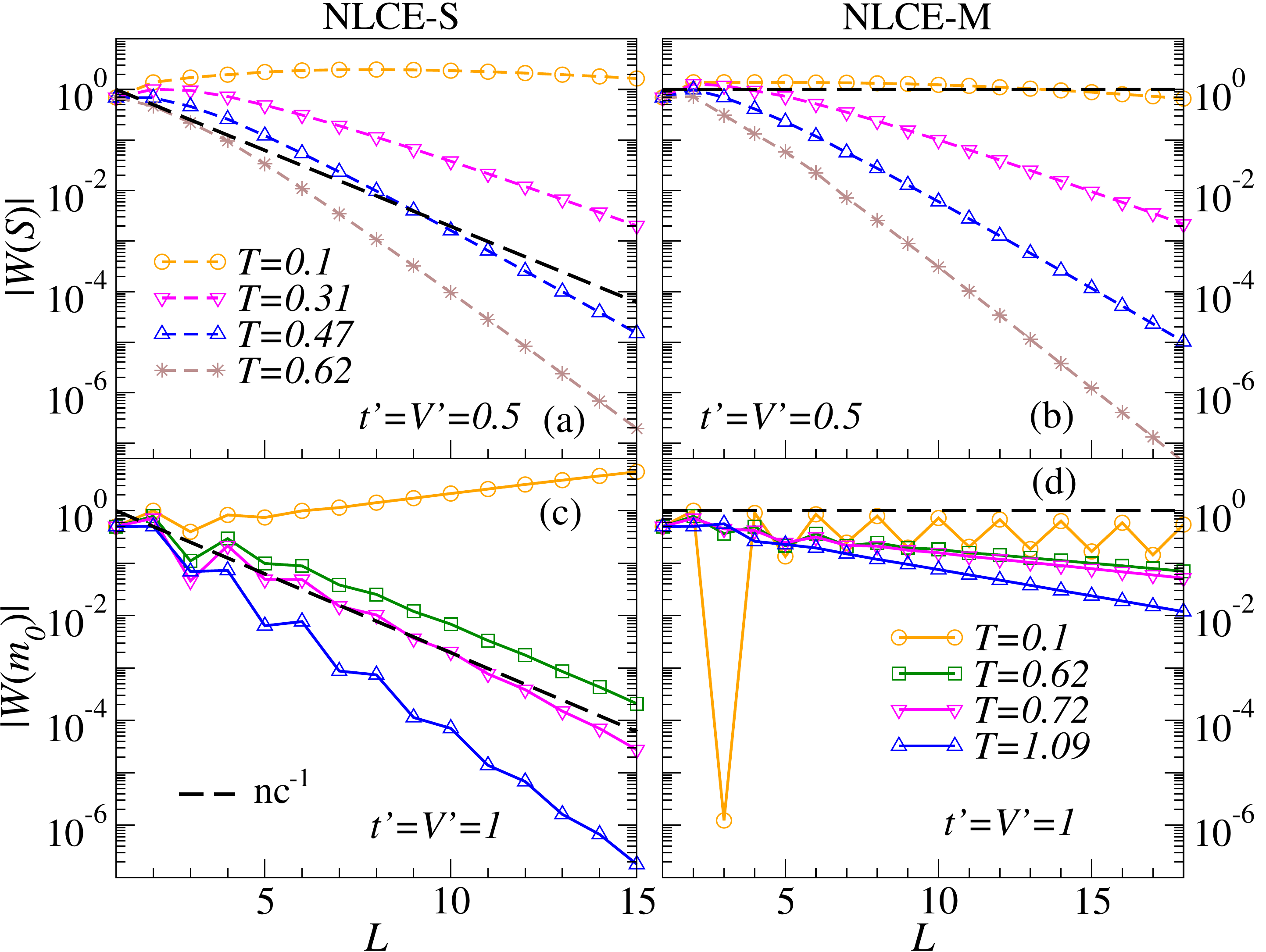}
\caption{Absolute value of the average of the cluster weights for the site-based [(a) and (c)] and the maximally connected [(b) and (d)] NLCEs vs the number of sites $L$ in the clusters, for the grand canonical ensemble at various $T$. We also show the inverse of the number of clusters vs $L$. (a) and (b) Average weights of the entropy. (c) and (d) Average weights of the  zero momentum mode occupation. For all cases, $t=V=1$.}\label{TH_weights}
\end{figure}

\begin{figure}[!t]
\includegraphics[width=.49\textwidth]{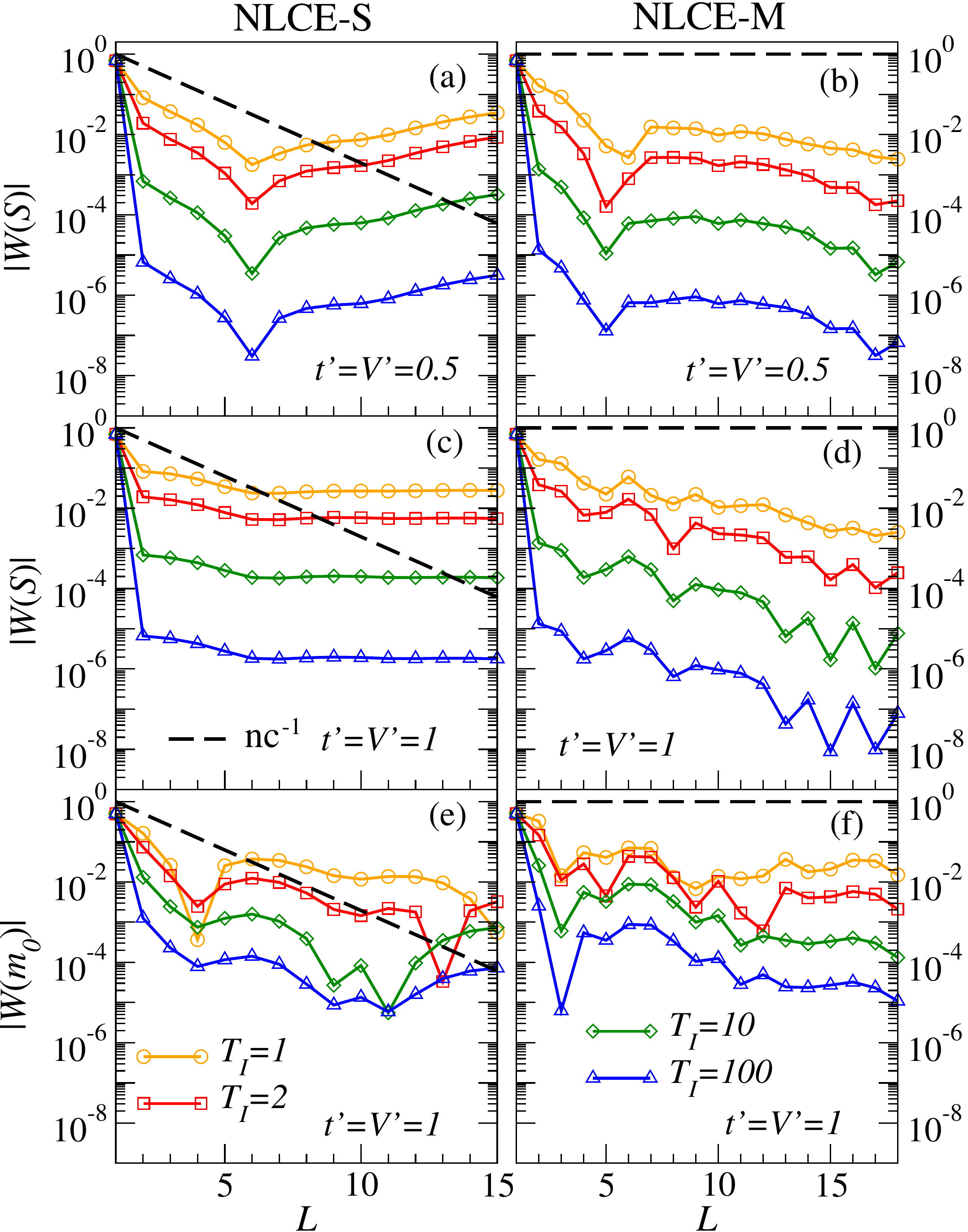}
\caption{The average weights in the diagonal ensemble after quantum quenches for site-based [(a),(c),and (e)] and maximally connected [(b),(d) and (f)] NLCEs. The temperatures reported in this case are the temperatures before the quench. For all the quenches, $t_I=0.5$, $V_I=1.5$, $t_I'=V_I'=0$, and $t=V=1$.}\label{DE_weights}
\end{figure}

In Fig.~\ref{TH_weights}, we plot the absolute value of the average weights in the grand canonical ensemble (at various temperatures) for two observables ($S$ and $m_0$), as well as the inverse of the number of clusters $\text{nc}^{-1}$, as a function of the cluster sizes $L$. The site-based NLCE can converge only if the average weight decreases more rapidly than $\text{nc}^{-1}$. In Figs.~\ref{TH_weights}(a) and \ref{TH_weights}(c), one can see that this happens for all $L$ only at the highest temperature shown. At the second highest temperature, the average weights decrease more slowly than $\text{nc}^{-1}$ for small $L$ and then more rapidly for large $L$ (hence the importance of calculating the NLCE up to largest possible cluster sizes). For the third highest temperature, the average weights decrease more slowly than $\text{nc}^{-1}$. For $T=0.1$, the lowest temperature shown, the average weights decrease only for the largest clusters for $S$ and increase for $m_0$.

For the maximally connected NLCE [Figs.~\ref{TH_weights}(b) and \ref{TH_weights}(d)], $\text{nc}^{-1}=1$ and the average weights can be seen to decrease exponentially fast with increasing $L$ for all but the lowest temperature (ensuring the convergence of this NLCE). We note that the average weight of $m_0$ in the site-based NLCE decreases much more rapidly than in the maximally connected one at high temperatures. This helps understand why the former NLCE could converge faster than the latter one in Fig.~\ref{energy_ns_GE}(c), and why the site-based NLCE can be used to complement the maximally connected one to study 1D lattice systems in thermal equilibrium.

The behavior of the average weights of observables in the diagonal ensemble after quantum quenches [Fig.~\ref{DE_weights}] is fundamentally different from that observed for systems in thermal equilibrium. Both, in the site-based and maximally connected NLCEs, the rate at which the weights decrease with $L$ does not appear to depend significantly on $T_I$, or the effective final $T$ (see Fig.~\ref{Ti_Tf} for the relation between the two), as opposed to the behavior seen in Fig.~\ref{TH_weights}. All the effect that those temperatures seem to have is to offset the decay of the weights at large $L$. In addition, while the weights in the maximally connected NLCE [see Figs.~\ref{DE_weights}(b),~\ref{DE_weights}(d), and~\ref{DE_weights}(f)] exhibit a decrease that is consistent with an exponential for large $L$ [except for the lowest temperatures in Fig.~\ref{DE_weights}(f)], the weights in the site-based NLCE exhibit an erratic behavior [see Figs.~\ref{DE_weights}(a),~\ref{DE_weights}(c), and~\ref{DE_weights}(e)] and fail to decrease faster than $\text{nc}^{-1}$. This explains why the site-based NLCE diverges with increasing the order of the expansion for all temperatures and cluster sizes considered. What determines the rate of decay of the weights in the diagonal ensemble after quantum quenches, and why their behavior is fundamentally different from the one in the grand canonical ensemble, is a topic that we will explore in future studies.

\section{Summary}\label{conclusion_sec}

In summary, we presented a comprehensive analysis of the performance of maximally connected and site-based NLCEs when applied to the study of hard-core boson systems in 1D lattices with nearest and next-nearest couplings in thermal equilibrium and after quantum quenches. We compared the results of the NLCE calculations with those of exact diagonalization in chains with periodic boundary conditions. Overall, we find that the maximally connected NLCE is the most efficient and provides the most accurate results in most regimes and for most observables studied. Only for some observables, such as the occupation of the zero momentum mode, in systems in thermal equilibrium at high temperatures, did we find that the site-based NLCE exhibits a faster convergence than the maximally connected one. We showed that when the NLCEs fail to converge, resummations can be used to accelerate convergence and extend the regime of applicability of NLCEs to lower temperatures.

We showed that, for observables that are not conserved during the quench, the convergence of the maximally connected NLCE with increasing cluster sizes is slower in the diagonal ensemble than in the grand canonical one. On the other hand, the site-based expansion was found to diverge with increasing cluster size for all initial temperatures and quenches considered. We argued that this is a result of the qualitatively different behavior between the weights of observables in the diagonal and the grand canonical ensembles (in the clusters considered here). Understanding the origin of those differences, as well as what determines the rate of decay of the weights in the diagonal ensemble, is left for future work.

\begin{acknowledgements}

This work was supported by the U.S. Office of Naval Research, Grant No.~N00014-14-1-0540. The computations were done at the Institute for CyberScience at Penn State.

\end{acknowledgements}

\bibliographystyle{apsrev4-1}

\bibliography{Reference}

\end{document}